\documentclass[conference]{IEEEtran}
\usepackage{graphicx}
\usepackage{array}
\newcolumntype{P}[1]{>{\centering\arraybackslash}p{#1}}
\newcolumntype{M}[1]{>{\centering\arraybackslash}m{#1}}
\ifCLASSINFOpdf
\else
\fi

\usepackage{fancyhdr}
\begin{document}

%
\title{And the Winner is ...: Bayesian Twitter-based Prediction on 2016 U.S. Presidential Election}

\author{\IEEEauthorblockN{Elvyna Tunggawan}
\IEEEauthorblockA{Information System Department\\
Universitas Multimedia Nusantara\\
Serpong, Indonesia 15811\\
Email: thoeng.elvyna@student.umn.ac.id}
\and
\IEEEauthorblockN{Yustinus Eko Soelistio}
\IEEEauthorblockA{Information System Department\\
Universitas Multimedia Nusantara\\
Serpong, Indonesia 15811\\
Email: yustinus.eko@umn.ac.id}}


%


\maketitle

\begin{abstract}
This paper describes a Naive-Bayesian predictive model for 2016 U.S. Presidential Election based on Twitter data. We use 33,708 tweets gathered since December 16, 2015 until February 29, 2016. We introduce a simpler data preprocessing method to label the data and train the model. The model achieves 95.8\% accuracy on \textit{10-fold cross validation} and predicts Ted Cruz and Bernie Sanders as Republican and Democratic nominee respectively. It achieves a comparable result to those in its competitor methods.

\end{abstract}


%
\IEEEpeerreviewmaketitle

\section{Introduction}
Presidential election is an important moment for every country, including the United States. Their economic policies, which are set by the government, affect the economy of other countries \cite{stokes}. On 2016 U.S. Presidential Election, Republican and Democratic candidates use Twitter as their campaign media. Previous researches have predicted the outcome of U.S. presidential election using Twitter \cite{wang, jahanbakhsh, oconnor, shi}. Some of them proved that Twitter data can complement or even predict the poll results. 
This follows the increasing improvement in the text mining researches \cite{soelistio, bermingham, talbot, gamallo, bakliwal}.

Some of the most recent studies are \cite{oconnor, jahanbakhsh, wang, chin}. Below we discuss these three recent studies and explain how our study relates to theirs. The first study is done by \cite{oconnor}, which analyzed the sentiment on 2008 U.S. Presidential Candidates by calculating sentiment ratio using moving average. They counted the sentiment value for Obama and McCain based on number of the positive and negative words stated on each tweet. The tweets were gathered during 2008-2009, whereas the positive and negative words were acquired from \textit{OpinionFinder}. They found that the comparison between sentiment on tweets and polls were complex since people might choose "Obama", "McCain", "have not decided", "not going to vote", or any independent candidate on the polls.

The second study predicted the outcome of 2012 U.S. Presidential Election polls using Naive Bayesian models \cite{jahanbakhsh}. They collected over 32 million tweets from September 29 until November 16, 2012. They used \textit{Tweepy} and set keywords for each candidate to collect the tweets, such as \textit{mitt romney}, \textit{barack obama}, \textit{us election}. The collected tweets passed some preprocessing stages: (1) URL, \textit{mentions}, \textit{hashtags}, \textit{RT}, and stop words removal; (2) tokenization; and (3) additional \textit{not\_} for negation. They analyzed 10,000 randomly selected tweets which only contain a candidate name. The analysis results were compared to \textit{Huffington Post}'s polls and they found that Obama's popularity on Twitter represented the polls result. This research didn't use tweets with two or more candidate names since it requires more complex preprocessing methods.

The third study built a system for real-time sentiment analysis on 2012 U.S. Presidential Election to show public opinion about each candidate on Twitter \cite{wang}. They collected tweets for each candidates using \textit{Gnip Power Track} since October 12, 2012 and tokenized them. The tweets were labeled by around 800 \textit{turkers} on \textit{Amazon Mechanical Turk} (AMT). They trained a \textit{Naive Bayes Classifier} using 17,000 tweets which consists of 4 classes: (1) positive; (2) negative; (3) neutral; and (4) unsure. It achieved 59\% accuracy, which is the best performance achieved in the three recent studies. They visualized the sentiment on a dashboard and calculated the trending words using TF-IDF.

As far as we know, there is not any research about prediction on 2016 U.S. Presidential Election yet. Previous researches either set the sentiment of a tweet directly based on a subjectivity lexicon \cite{oconnor} or preprocessed the tweet using a complex preprocessing method \cite{wang, jahanbakhsh}. \cite{jahanbakhsh} not only removed URLs, \textit{mentions}, \textit{retweets}, \textit{hashtags}, numbers and stop words; but also tokenized the tweets and added \textit{not\_} on negative words. \cite{wang} tokenized the tweets and separated URLs, \textit{emoticons}, phone numbers, HTML tags, \textit{mentions}, \textit{hashtags}, fraction or decimals, and symbol or Unicode character repetition. This research analyzes sentiment on tweets about 2016 U.S. Presidential candidates. We will build a Naive Bayesian predictive model for each candidate and compare the prediction with \textit{RealClearPolitics.com}. We expect to have a correct prediction on the leading candidates for Democratic and Republican Party. We prove that using a simpler preprocessing method can still have comparable performance to the best performing recent study \cite{wang}.

We explain our data preparation methods in the next section. It is followed by our research methodology in Section III. We present our results in Section IV, which is followed by discussion and conclusion in Section V and VI.

\section{Data Preparation}

\subsection{Data Collection}
We gathered 371,264 tweets using \textit{Twitter Streaming API} on \textit{Tweepy} \cite{jahanbakhsh} since December 16, 2015 until February 29, 2016. We use \emph{\#Election2016} as the search keyword since it is the official \textit{hashtag} used during 2016 U.S. Presidential Election cycle and it covers conversations about all candidates. We separate the tweets per period, which is seven days. Figure 1 shows tweets frequency distribution, with the average of 37,126.4 tweets per period and standard deviation 27,823.82 tweets. Data collection from January 20 to January 26, 2016 are limited due to resource limitation. The data are saved as JSON files.

\begin{figure}[!t]
\centering
\includegraphics[scale=0.5]{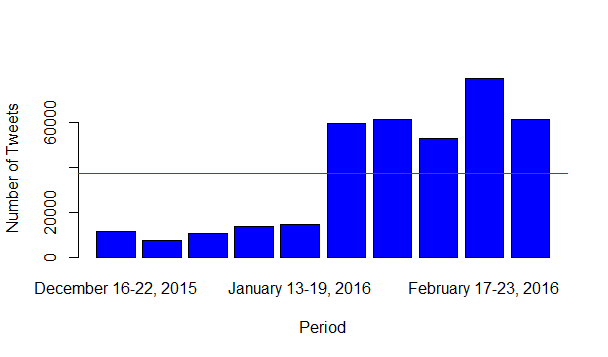}
\caption{Collected Tweets Distribution ($\mu=37,126.4; \sigma=27,823.82$)}
\label{coll_tweet_dist}
\end{figure}

Each line of the JSON files represents a tweet, which consists of 26 main attributes, such as \textit{created\_at}, \textit{ID}, \textit{text}, \textit{retweet\_count}, and \textit{lang}. We only use the contents of \textit{created\_at} and \textit{text} attributes since this research focuses on the sentiment toward the candidates in a particular time, not including the geographic location and other information. The collected tweets are mainly written in English. We publish the raw and preprocessed tweets upon request for future use. The data are available for research use by email.

\subsection{Data Preprocessing}
We preprocess the data by: (1) removing URLs and pictures, also (2) by filtering tweets which have candidates' name. \textit{Hashtags}, \textit{mentions}, and \textit{retweets} are not removed in order to maintain the original meaning of a tweet. We only save tweets which have passed the two requirements such as in Table 1. The first example shows no change in the tweet's content, since there isn't any URLs or pictures, and it contains a candidate's name: Bernie Sanders. The second example shows a removed tweet, which doesn't contain any candidates' name. The preprocessing stage changes the third tweet's contents. It removes the URLs and still keeps the tweet because it contains "Hillary Clinton" and "Donald Trump". The preprocessing stage removes 41\% of the data (Figure 2).

\begin{table}[!t]
\renewcommand{\arraystretch}{1.3}
\caption{Tweets Example on Preprocessing Stage}
\label{table_preprocess}
\centering
\begin{tabular}{|c|p{3cm}|p{3cm}|}
\hline
No & Before & After\\
\hline
1 & I'm shocked, SHOCKED to hear that @thecjpearson is now a \#BernieSanders supporter. Never saw that coming \#Election2016 & I'm shocked, SHOCKED to hear that @thecjpearson is now a \#BernieSanders supporter. Never saw that coming \#Election2016\\
\hline
2 & Stats from @actblue show a clear picture of the small-donor revolution. https://t.co/ddw893OuLF \#Election2016 https://t.co/56J1fX4YPQ & -\\
\hline
3 & Hillary Clinton \&amp; Donald Trump Make Political History. But Not In A Good Way https://t.co/SxUxjHjaa3 via @POStqia \#election2016 \#polling & Hillary Clinton \&amp; Donald Trump Make Political History. But Not In A Good Way via @POStqia \#election2016 \#polling\\
\hline
\end{tabular}
\end{table}

\begin{figure}[!t]
\centering
\includegraphics[scale=0.5]{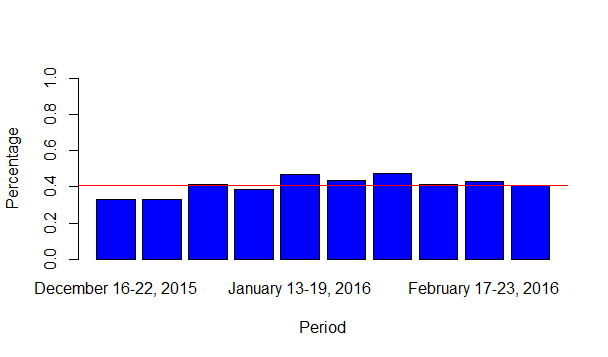}
\caption{Removed Tweets on Preprocessing Stage ($\mu=40.87\%; \sigma=4.98\%$)}
\label{removed_tweet_dist}
\end{figure}

\subsection{Data Labeling}
The preprocessed tweets are labeled manually by 11 annotators who understand English. All annotators are given either grade as part of their coursework or souvenirs for their work. The given label consists of the intended candidate and the sentiment. The annotators interpret the tweet and decide whom the tweet relates to. If they think the tweets does not relate to particular candidate nor understand the content, they can choose "not clear" as the label. Otherwise, they can relate it to one candidate and label it as positive or negative. We divide the tweets and annotators into three groups (Table II). They label as many tweets as they can since January 24 until April 16, 2016.

The validity of the label is determined by means of majority rule \cite{dasgupta}. Each tweet is distributed to three or five annotators and it is valid when there is a label which occurs the most. As the final data preparation step, we remove all "not clear" labeled tweets. Figure 3 shows the distribution of tweet labels. Most tweets are related to Bernie Sanders, Donald Trump, and Hillary Clinton.


\begin{table}[!t]
\renewcommand{\arraystretch}{1.3}
\caption{Tweets Distribution on Labeling Stage}
\label{table_label_dist}
\centering
\begin{tabular}{|P{4.5cm}|c|}
\hline
Tweet Period & Number of Annotators\\
\hline
December 16,2015-January 19, 2016 & 5\\
\hline
January 27-February 2, 2016 & 3\\
\hline
February 3-9, 2016 & 3\\
\hline
\end{tabular}
\end{table}


\begin{figure}[!t]
\centering
\includegraphics[scale=0.7]{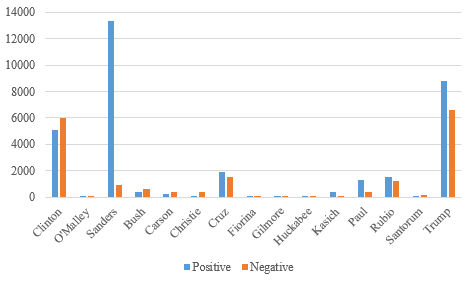}
\caption{Sentiment Distribution by Candidates}
\label{candidate_sentiment_dist}
\end{figure}

\section{Methodology}
The presidential nominees are predicted by finding candidates with the most predicted positive sentiment. The sentiments are predicted using Bayesian model. This section describes: (1) the model training, (2) model accuracy test, and (3) prediction accuracy test.

\subsection{Model Training}
Our models are trained using \emph{Naive Bayes Classifier}. We have one model representing each candidate, consequently we have 15 trained models. We use \emph{nltk.classify} module on \textit{Natural Language Toolkit} library on \textit{Python}. We use the labeled data gathered since December 16, 2015 until February 2, 2016 as training data to our models. The rest of our labeled data will be used to evaluate the models.

\subsection{Model Accuracy Test}
Our models' accuracy is tested using \emph{10-fold cross validation}. Model validation is done using \textit{scikit-learn} library. The accuracy is calculated by checking the confusion matrix \cite{metz, kamber} and its $ F_1 $ score \cite{witten}.

On some folds, the models predict the sentiment in extreme value (i.e. only have positive or negative outcomes). Due to these cases, we can not calculate $ F_1 $ score of Chris Christie's model. The average accuracy and $ F_1 $ score are 95.8\% and 0.96 respectively.

$ F_1 $ score only measures how well the model works on predicting positive sentiment, so we propose a modified $ F_1 $ score ($\sim F_1 $) by reversing the formula. $ \sim F_1 $ score shows how well the model predicts negative sentiment. 

\begin{equation}
\sim F_1=2\times\frac{\frac{TN}{TN+FN}\times\frac{TN}{TN+FP}}{\frac{TN}{TN+FN}+\frac{TN}{TN+FP}}
\end{equation}

\subsection{Prediction Accuracy Test}
The models use tweets gathered from February 3 until 9, 2016 as the prediction input. The prediction follows two steps: (1) we calculate the positive sentiment from tweets and consider the number of positive sentiment as the likelihood of a candidate to be the nominee, and (2) we sort the candidates by number of their positive sentiment. The ranks are compared to the poll results on \emph{RealClearPolitics.com}. We calculate the error rate (\textit{E}) by dividing the difference of the poll rank with our predicted rank with number of candidates ($e_i$).


\begin{equation}
e_i = |Po_i - Pre_i|
\end{equation}
\begin{equation}
E=\frac{\sum_i^n e_i}{n}
\end{equation}

where $1\leq\emph{i}\leq\emph{n}$ and \emph{n} equals the number of candidates. \textit{Po} and \textit{Pre} are the poll and prediction ranks associated with \emph{RealClearPolitics.com} and the model respectively.

\section{Results}

\subsection{Model Accuracy Test}
The models show good accuracy and $F_1$ score (Table III). It shows that the model can predict the test data almost perfectly (95.8\%) with slightly better result on positive sentiment than negative ones, which can be seen by the larger value of $F_1$ than $\sim F_1$.

The test results do not show exact effect of training data and the model accuracy. Models with smaller number of training data (e.g. Huckabee's, Santorum's) achieve higher accuracy than models with larger number of training data (e.g. Trump's, Clinton's), while the lowest accuracy is achieved by Kasich's, which is trained with small number of training data. The undefined value of $F_1$ and $\sim F_1$ scores on Christie's, Gilmore's, and Santorum's model shows extreme predictions on these models.


\begin{table}[!t]
\renewcommand{\arraystretch}{1.3}
\caption{Model Test Results}
\label{table_10_fold}
\centering
\begin{tabular}{|c|c|c|c|c|}
\hline
Candidate & Training Data & Accuracy & $ F_1 $ & $ \sim F_1 $\\
\hline
Clinton & 7,672 & 0.953 & 0.941 & 0.96\\
\hline
O'Malley & 141 & 0.964 & 0.977 & 0.89\\
\hline
Sanders & 9,224 & 0.982 & 0.99 & 0.894\\
\hline
Bush & 674 & 0.973 & 0.964 & 0.943\\
\hline
Carson & 529 & 0.915 & 0.882 & 0.932\\
\hline
Christie & 200 & 0.965 & - & 0.977\\
\hline
Cruz & 1,998 & 0.967 & 0.975 & 0.952\\
\hline
Fiorina & 83 & 0.965 & 0.955 & 0.969\\
\hline
Gilmore & 35 & 0.942 & 0.947 & -\\
\hline
Huckabee & 88 & 1.00 & 1.00 & 1.00\\
\hline
Kasich & 350 & 0.88 & 0.929 & 0.578\\
\hline
Paul & 1,008 & 0.958 & 0.977 & 0.767\\
\hline
Rubio & 1,491 & 0.97 & 0.971 & 0.968\\
\hline
Santorum & 102 & 0.981 & 0.98 & -\\
\hline
Trump & 10,113 & 0.95 & 0.945 & 0.954\\
\hline
\multicolumn{2}{|c|}{Average} & 0.958 & 0.96 & 0.906\\
\hline
\end{tabular}
\end{table}

\subsection{Prediction Accuracy Test}
We use tweets on February 3-9, 2016 as the input to our models, regarding to the specified candidate. We rank the prediction result by sorting the number of positive predictions on each candidate. On Democratic Party, Bernie Sanders leads the rank with 3,335 tweets, followed by Martin O'Malley (14 tweets) and Hillary Clinton (none). The prediction ranks on Republican Party are (1) Ted Cruz (1,432 tweets), (2) Marco Rubio (1,239 tweets), (3) Rand Paul (645 tweets), (4) Rick Santorum (186 tweets), (5) John Kasich (133 tweets), (6) Carly Fiorina (88 tweets), (7) Mike Huckabee (11 tweets), and (8) Jim Gilmore (5 tweets). The other Republican candidates do not have any positive prediction, so we place them at the bottom rank. 

\begin{table}[!t]
\renewcommand{\arraystretch}{1.3}
\caption{Prediction Error on Democratic Candidates}
\label{table_error_dem}
\centering
\begin{tabular}{|c|c|c|c|}
\hline
Candidate & Prediction & Poll & $ e_i $\\
\hline
Clinton & 3 & 1 & 2\\
\hline
O'Malley & 2 & 3 & 1\\
\hline
Sanders & 1 & 2 & 1\\
\hline
\multicolumn{3}{|c|}{Error Rate} & 1.33\\
\hline
\end{tabular}
\end{table}

\begin{table}[!t]
\renewcommand{\arraystretch}{1.3}
\caption{Prediction Error on Republican Candidates}
\label{table_error_republican}
\centering
\begin{tabular}{|c|c|c|c|c|}
\hline
Candidate & Prediction & Prediction (Adjusted) & Poll & $ e_i $\\
\hline
Bush & 9 & 8 & 5 & 3\\
\hline
Carson & 9 & 8 & 8 & 0\\
\hline
Christie & 9 & 8 & 8 & 0\\
\hline
Cruz & 1 & 1 & 2 & 1\\
\hline
Fiorina & 6 & 6 & 7 & 1\\
\hline
Gilmore & 8 & 8 & 8 & 0\\
\hline
Huckabee & 7 & 7 & 8 & 1\\
\hline
Kasich & 5 & 5 & 6 & 1\\
\hline
Paul & 3 & 3 & 8 & 5\\
\hline
Rubio & 2 & 2 & 3 & 1\\
\hline
Santorum & 4 & 4 & 4 & 0\\
\hline
Trump & 9 & 8 & 1 & 7\\
\hline
\multicolumn{4}{|c|}{Error Rate} & 1.67\\
\hline
\end{tabular}
\end{table}

Our model prediction ranks from 1 to 9 and it differs from the poll's (rank 1 to 8). Before we do the comparison, we adjust the prediction ranks in order to make an equal range. We move Jeb Bush, Ben Carson, Chris Christie, and Donald Trump, who are formerly on the 9th rank, to the 8th rank. We compare the prediction ranks with the poll and calculate the error rate. Our model gets 1.33 error of 2 remaining Democratic candidates, which we consider not good. Our model performs better on predicting Republican candidates, which achieves 1.67 error of 7 remaining candidates (see Table IV and V).

Overall prediction accuracy can be calculated by subtracting one with the average result of error rate division on each party by number of its remaining candidates. We achieve 0.548 prediction accuracy, which is not good enough\cite{wang}. The model accuracy is mainly affected by the large error rate on Democratic candidates (1.33 from 2 candidates).


\section{Discussion}
Using simple preprocessed data, our Naive Bayesian model successfully achieves 95.8\% accuracy on \emph{10-fold cross validation} and gets 54.8\% accuracy on predicting the poll result.  The model predicts Ted Cruz and Bernie Sanders as the nominee of Republican and Democratic Party respectively. Based on the positive predictions, it predicts that Bernie Sanders will be elected as the 2016 U.S. President.

Although it has 95.8\% accuracy during the model test, the model's prediction does not represent the poll. Table III shows that model's accuracy is not dependent of its number of training data. Model with less training data (e.g. Mike Huckabee's) can perform perfectly during the model test and only misses a rank on the prediction, whereas model with more training data (e.g. Donald Trump's) can have worse performance. 

To see how the model accuracy is affected by number of training data, we train more models for each candidate using \textit{n} first tweets and use them to predict the next 4000 tweets' sentiment (see Figure 4). Bernie Sanders' and Donald Trump's models have the most consistent accuracy on predicting the sentiment. Models with less training data (e.g. Martin O'Malley, Jim Gilmore, Mike Huckabee) tend to have fluctuating accuracy. The models which are trained using 1,000 first tweets have 55.85\% of average accuracy and 26.49\% of standard deviation, whereas the models which are trained using 33,000 first tweets have slightly different accuracy: 65.75\% of average accuracy and 27.79\% of standard deviation. This shows that the number of training data does not affect the overall model accuracy.

Our model might not represent the poll, but the election is still ongoing and we do not know which candidate will become the next U.S. President. Hence, there is possibility that the predicted nominees become the next U.S. President. Otherwise, Twitter might not be used to predict the actual polls \cite{gayo}.

\begin{figure}[!t]
\centering
\includegraphics[scale=0.55]{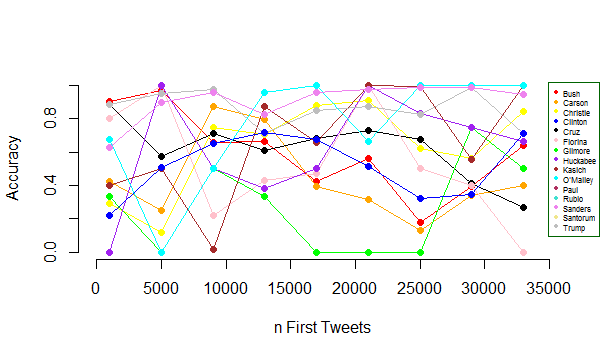}
\caption{Model Test using \textit{n} First Tweets as Training and 4000 Next Tweets as Test Data}
\label{partial_model_n_test_data}
\end{figure}

\section{Conclusion}
We built Naive Bayesian predictive models for 2016 U.S. Presidential Election. We use the official hashtag and simple preprocessing method to prepare the data without modifying its meaning. Our model achieves 95.8\% accuracy during the model test and predicts the poll with 54.8\% accuracy. The model predicts that Bernie Sanders and Ted Cruz will become the nominees of Democratic and Republican Party respectively, and the election will be won by Bernie Sanders.






%

\end{document}